\documentclass[journal]{IEEEtran}

\usepackage{amsmath,amsfonts}
\usepackage{algorithmic}
\usepackage{array}
\usepackage[caption=false,font=normalsize,labelfont=sf,textfont=sf]{subfig}
\usepackage{textcomp}
\usepackage{stfloats}
\usepackage{url}
\usepackage{verbatim}
\usepackage{graphicx}
\usepackage{graphicx}
\usepackage{ascmac}

\usepackage{bm}
\usepackage{amsmath}
\usepackage{amssymb}
\usepackage{amsfonts}
\usepackage{color}
\usepackage{wrapfig}
\usepackage{url}
\usepackage{cite}
\usepackage{siunitx}
\usepackage{mathtools}

\usepackage{diagbox}
\usepackage{enumerate}

\usepackage{comment}

\def\BibTeX{{\rm B\kern-.05em{\sc i\kern-.025em b}\kern-.08em
    T\kern-.1667em\lower.7ex\hbox{E}\kern-.125emX}}
\usepackage{balance}

\begin{document}

\title{Stability analysis for circulant structured multi-agent molecular communication systems}

\author{Taishi~Kotsuka,~
        Yutaka~Hori,~\IEEEmembership{Member,~IEEE}
        
\thanks{This work was supported in part by JSPS KAKENHI Grant Numbers JP21H05889 and JP22J10554, and in part by JST SPRING Grant Number JPMJSP2123.}
\thanks{T. Kotsuka and Y. Hori are with the Department
of Applied Physics and Physico-Informatics, Keio University, Kanagawa 223-8522 Japan. Correspondence should be addressed to Y. Hori (email: tkotsuka@keio.jp; yhori@appi.keio.ac.jp).}

}

\maketitle

\begin{abstract}
In this paper, we introduce the system theoretic model for the multi-agent MC systems represented by multi-input and multi-output (MIMO) systems using the transfer functions, and then propose a method to analyze the stability for the special case of the circulant structured multi-agent MC systems. The proposed method decomposes the MIMO MC system into multiple single-input and single-output (SISO) systems, which facilitates to analyze of the stability of the large-scale multi-agent MC system. Finally, we demonstrate the proposed method to analyze the stability of a specific MC system.
\end{abstract}


\IEEEpeerreviewmaketitle

\section{Introduction}

Recently, many studies for molecular communication (MC) systems were conducted to control the behavior of dispersed nanorobot populations toward engineering applications such as targeted drug delivery \cite{tatsuya2018molecular,Bi2021,Soldner2020,Femminella2015,Gao2014}. In general MC systems, the dispersed nanorobots transfer signals to each other via MC channels to achieve cooperative performance, and the fundamental properties of such MC channels were analyzed based on diffusion equation \cite{Pierobon2010,Chude-Okonkwo2015,Huang2021,kotsuka2022frequency}. 
One of the key roles of MC in the control of the multi-agent system is to stabilize and synchronize the reactions inside the individual nanorobot. 
To control the behavior of nanorobot populations, the dynamical reaction system in each nanorobot needs to be stabilized via the MC channels. 
To this end, it is crucial to model the multi-agent MC systems that incorporates the dynamics of reactions in the nanorobots and develop tools for stability analysis.

\smallskip
\par
There are several studies of multi-agent MC systems considering the reaction in the nanorobots based on the reaction-diffusion (RD) equation \cite{Kotsuka2022,yutaka2015coordinated,Hsia2012}. However, RD-based MC models assume that the distance between the nanorobots are close enough to ignore disruption of the signal by the MC channel, and thus, the model is not suitable for the analysis and design of the MC systems composed of a population of distributed nanorobots for applications in vivo such as drug delivery.
This limitation was somewhat relaxed by introducing a system theoretic model of multi-agent dynamical nanorobots that incorporates a detailed model of reactions and diffusion by the authors' group \cite{Hara2021}. However, the previously proposed stability analysis method \cite{Hara2021} was applicable only to the MC systems with two nanorobots. This motivates us to develop a more versatile stability analysis method for large-scale multi-agent MC systems for broader engineering applications of MC systems.

\smallskip
\par
In this paper, we introduce a system theoretic model of multi-agent MC systems that incorporates the transfer functions of bidirectional MC channels and local reactions inside nanorobots, and then, propose a stability analysis method for a special case of the circulant structured homogeneous multi-agent MC systems. 
The proposed method decomposes the transfer function matrix of the MC system into multiple single-input and single-output (SISO) systems in the same spirit as the stability analysis for linear systems with generalized frequency variables \cite{hara2009LTI}. This decomposition allows for the stability analysis of the large-scale multi-agent MC system by combining simple analysis techniques for SISO systems. Finally, we demonstrate the proposed method to analyze the stability of a specific MC system.

\section{Multi-agent MC systems}
\label{sec:model}

\begin{figure}
    \centering
    \includegraphics[width=0.99\linewidth]{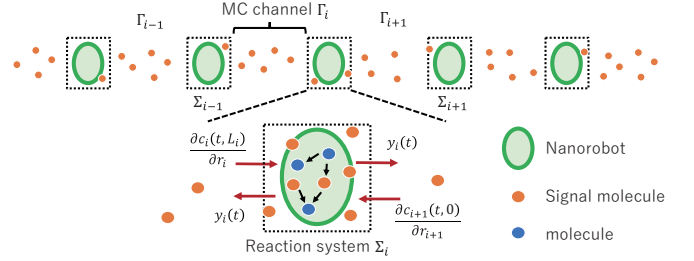}
    \caption{The multi-agent MC system}
    \label{fig:mcsystems}
\end{figure}

We consider one-dimensional multi-agent MC systems consisting of $n$ nanorobots and $(n+1)$ MC channels shown in Fig. \ref{fig:mcsystems}. 
Let $\Sigma_i$ denote the $i$-th reaction system that captures the reaction in the $i$-th nanorobot and the diffusion of the signal molecules, and $\Gamma_i = [0,L_i]$ denote the $i$-th MC channel between the reaction systems or between the reaction system and the left/right boundaries of the MC system with $L_i$ being the communication distance.

\par
\smallskip
We denote the location and the concentration of the signal molecules in the MC channel $\Gamma_i$ by $r_i$ and $c_i(t,r_i)$, respectively. The dynamics of the concentration $c_i(t,r_i)$ can be modeled by the diffusion equation 
\begin{equation}
    \frac{\partial c_i(t,r_i)}{\partial t} = \mu\frac{\partial^2c_i(t,r_i)}{\partial r^2_i},
    \label{eq:diffusion}
\end{equation}
where $\mu$ is the diffusion coefficient. The boundary conditions at $r_i=0$ and $r_i=L_i$ are Dirichlet boundary conditions as $c_i(t,0)=y_{i-1}(t)$ and $c(t,L_i)=y_i(t)$ for $i=2,3,\cdots,n$, respectively. 
Dirichlet or Neumann boundary conditions can be applied to the left end boundary of $\Gamma_1$ and the right end boundary of $\Gamma_{n+1}$ depending on the situations. 
The initial condition is
\begin{equation}
    c_i(0,r_i) = 0; \ \forall r_i \in [0,L_i].
\end{equation}

\par
\smallskip
For many practical examples, the $i$-th reaction system $\Sigma_i$ can be modeled by non-linear state-space models as
\begin{eqnarray}
    \frac{d\bm{x}(t)}{dt} &=& \bm{f}(\bm{x}) + \bm{B} w_{i}(t),\nonumber\\
    y_{i}(t) &=& \bm{C}\bm{x}(t),\label{eq:Rsystem}
\end{eqnarray}
where the state $\bm{x}(t)\in\mathbb{R}^m$ is the concentrations of the molecules associated with reactions occurring inside and outside of the nanorobot, and the $m$-th entry of $\bm{x}(t)$ represents the concentration of the signal molecule outside of the nanorobot. The vector function $\bm{f}(\cdot)$ represents the dynamics of the reactions in the nanorobot and the membrane transport, $\bm{B}=[0,\cdots,0,1]^T\in\mathbb{R}^{m}$ and $\bm{C}=[0,\cdots,0,1]\in\mathbb{R}^{m}$ are the input and output vectors, respectively. 
The variable $w_i(t)$ is the input fluxes from the MC channels $\Gamma_i$ and $\Gamma_{i+1}$ to the $i$-th reaction system $\Sigma_i$ represented by 
\begin{equation}
w_{i}(t) = \frac{\mu}{\Delta r_i}\left(\frac{\partial c_{i+1}(t,0)}{\partial r_{i+1}} - \frac{\partial c_i(t,L_i)}{\partial r_i}\right),
\end{equation}
where $\Delta r_i$ is the size of the reaction system $\Sigma_i$. The output $y_i(t)$ is the concentration of the signal molecule outside of the nanorobot .

\par
\smallskip
The dynamical multi-agent MC system can synchronously converge to a spatially homogeneous equilibrium point $\bm{x}_1(t) = \bm{x}_2(t) = \cdots = \bm{x}_n(t)$ at steady state if the reactions are appropriately designed. In what follows, we are interested in the local stability analysis of the multi-agent MC system. Specifically, we first express the multi-agent MC system as a multi-input multi-output (MIMO) dynamical system using the transfer functions derived from the diffusion equation (\ref{eq:diffusion}) and the state-space model (\ref{eq:Rsystem}). We then show that the circulant structured MIMO MC system can be decomposed into multiple single-input single-output (SISO) systems under certain conditions to facilitate the stability analysis.

\section{Stability analysis for multi-agent MC systems}
\label{sec:stability}

In this section, we first introduce the MIMO representation of the MC system. We then show that the stability analysis of the $n$-dimensional MIMO MC system can be reduced to that of $n$ SISO systems for a specific class of the MC channel.

\subsection{System theoretic model}

\begin{figure}
    \centering
    \includegraphics[width=0.99\linewidth]{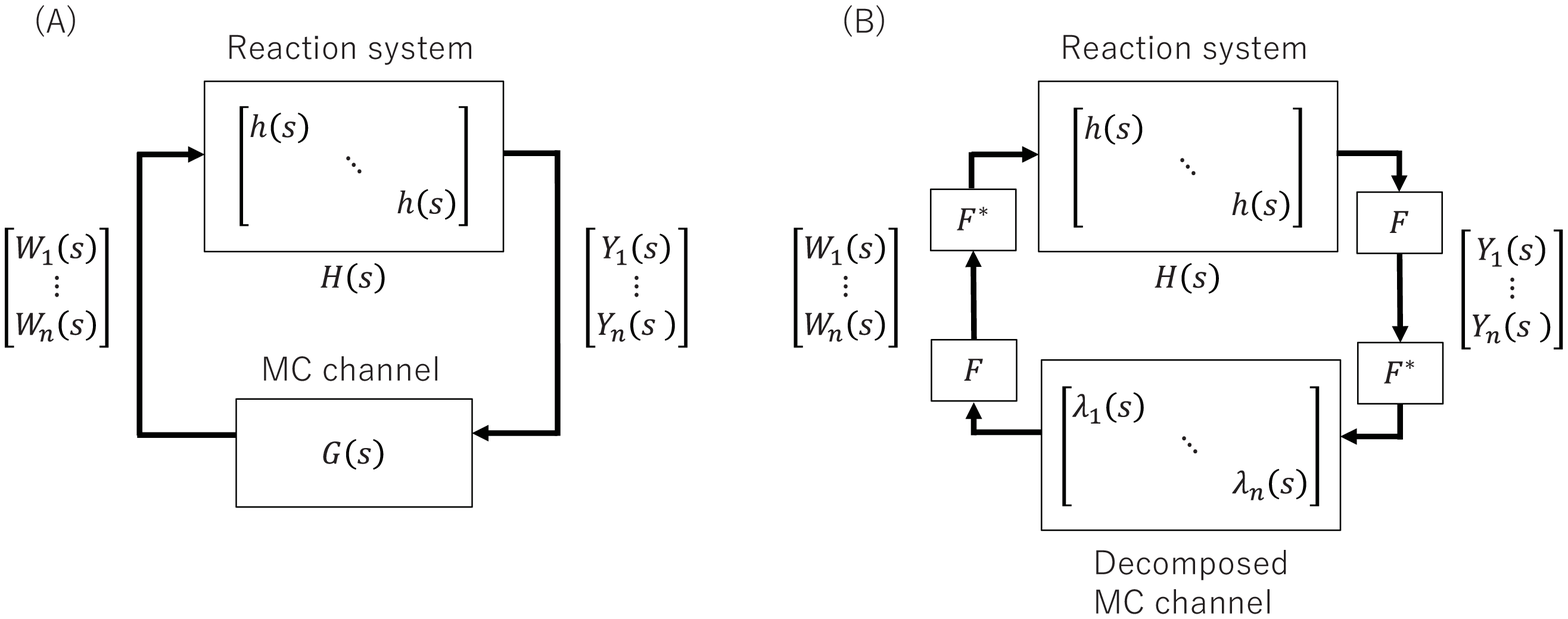}
    \caption{(A) The MIMO representation of the multi-agent MC system. (B) The decomposed multi-agent MC system.}
    \label{fig:mimo}
\end{figure}

The MIMO representation of the MC system consisting of the MC channels $\Gamma_i$ and the reaction systems $\Sigma_i$ shown in Fig. \ref{fig:mimo} (A) can be expressed as 
\begin{equation}
\begin{split}
    \bm{Y}(s)
    &= H(s) \bm{W}(s),\\
    \bm{W}(s)
    &= G(s) \bm{Y}(s),
\end{split}\label{eq:mimo}
\end{equation}
where $\bm{Y}(s):=[Y_1(s),Y_2(s),\cdots,Y_n(s)]^T$ and $\bm{W}(s):=[W_1(s),W_2(s),\cdots,W_n(s)]^T$ with $Y_i(s)$ and $W_i(s)$ being the Laplace transform of $y_i(t)$ and $w_i(t)$ as $Y_i(s):=\mathcal{L}[y_i(t)]$ and $W_i(s):=\mathcal{L}[w_i(t)]$. The transfer function matrix $H(s)$ is the diagonal matrix representing the reaction systems, where the $(i,i)$-th entry is $h(s)=\bm{C}(sI-A)^{-1}\bm{B}$ and $A$ is the Jacobian matrix of $\bm{f(\bm{x})}$ at the homogeneous equilibrium. The transfer function matrix $G(s)$ is the symmetric matrix representing the MC channels defined by 
\begin{equation}
    G(s) = \left[\begin{matrix}
        g_{1}^b(s)&g_{1}^s(s)&0&\cdots&0&g_{p}^s(s)\\g_{1}^s(s)&g_{2}^t(s)&g_{2}^s(s)&\cdots&0&0\\0&g_{2}^s(s)&\ddots&\ddots&0&\vdots \\\vdots&\ddots&\ddots&\ddots&\ddots&0 \\0&\ddots&\ddots&\ddots&\ddots&g_{n-1}^s(s) \\g^s_{p}(s)&0&\cdots&0&g^s_{n-1}(s)&g^b_{n}(s)
    \end{matrix} \right],\label{eq:Gsp}
\end{equation}
where
\begin{eqnarray}
    g^t_i(s) &=& -\hat{s}\left(\frac{1}{\tanh{\left(L_{i}\hat{s} \right)}} + \frac{1}{\tanh{\left(L_{i+1}\hat{s} \right)}} \right),\label{eq:Gtan}\\
    g^s_i(s) &=& \frac{\hat{s}}{\sinh{\left(L_{i+1}\hat{s} \right)}}\label{eq:Gsin}
\end{eqnarray}
with $\hat{s}=\sqrt{s/\mu}$. The functions $g^b_1(s)$ and $g^b_n(s)$ are determined by the boundary conditions for the left end boundary of $\Gamma_1$ and the right end boundary of $\Gamma_n$, respectively. 
The function $g^s_p(s)$ represents the MC channel between the reaction systems $\Sigma_1$ and $\Sigma_n$ with distance $L_0$, which appears when the periodic boundary $c_1(t,0)=y_n(t)$ is applied as
\begin{align}
 g^s_p(s) =
\begin{cases}
\displaystyle 
\frac{\hat{s}}{\sinh{\left(L_{0}\hat{s} \right)}} & (\mathrm{Periodic\,boundary})\\
\displaystyle
0 & (\mathrm{Otherwise}).
\end{cases}
\label{eq:v-z-def}
\end{align}
The functions $g^t_i(s)$ and $g^s_i(s)$ are obtained by the Laplace transform of the diffusion equation (\ref{eq:diffusion}) with Dirichlet boundary conditions.

\smallskip
\par
\begin{figure}
    \centering
    \includegraphics[width=0.99\linewidth]{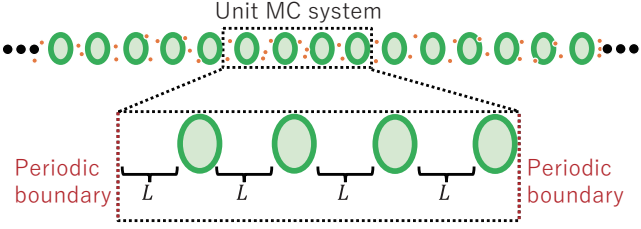}
    \caption{The multi-agent MC system in Example 1 and the unit MC system.}
    \label{fig:circ}
\end{figure}

\noindent
{\bf Example 1.} 
Consider an MC system, where countless homogeneous nanorobots are spaced with same distance $L$ as shown in Fig. \ref{fig:circ}. The MC system can be approximately modeled by the unit MC system composed of $n$ nanorobots with the periodic boundary $c_1(t,0)=y_n(t)$, which leads to $g^b_1(s)=g^t_1(s)$ and $g^b_n(s)=g^t_n(s)$. The reaction system is $H(s)=h(s)I$, where $h(s)=\bm{C}(sI-A)^{-1}\bm{B}$. The MC channel $G(s)$ of the unit MC system is modeled by a circulant transfer matrix
\begin{equation}
    G(s)=\mathrm{circ}(g^t(s),g^s(s),0,\cdots,0,g_p^s(s)),\label{eq:circulant}
\end{equation}
where $g^t(s)$ and $g^s(s)$ are defined by (\ref{eq:Gtan}) and (\ref{eq:Gsin}) with $L_i=L_{i+1}=L$, respectively.

\subsection{Stability condition for periodic MC systems}

We here show a necessary and sufficient stability condition for the closed-loop system (\ref{eq:mimo}) based on the characteristic equation. The characteristic polynomial $p(s)$ of the closed-loop system (\ref{eq:mimo}) is 
\begin{equation}
    p(s) := \det(I-H(s)G(s)).\label{eq:character}
\end{equation}
The closed-loop is asymptotically stable if and only if the real part of all the roots of the characteristic equation $p(s)=0$ are negative.

\par
\smallskip
The roots of the characteristic equation $p(s)$ are not easy to analyze since it involves the computation of the determinant. 
In what follows, we show a theorem that the stability analysis for the MIMO system (\ref{eq:mimo}) with the circulant matrix $G(s)$ can be reduced to that for $n$ SISO systems. 

\par
\medskip
\noindent
{\bf Theorem 1.}
Consider the MC system with the homogeneous reaction systems $h(s)$ and the MC channel (\ref{eq:circulant}) with the periodic boundary $c_1(t,0)=y_n(t)$. The closed-loop MC system is asymptotically stable if and only if the real part of all the roots of the characteristic equations
\begin{eqnarray}
    \hat{p}_i(s) := 1-h(s)\lambda_i(s) = 0\label{eq:chat}
\end{eqnarray}
are negative for all $i\,(i=1,2,\cdots,n)$, where 
\begin{equation}
    \lambda_i(s) = g^t_{i}(s) + g^s_i(s)\left(\alpha^{i-1} + \alpha^{(i-1)(n-1)} \right)
\end{equation}
with $\alpha = e^{-j2\pi/n}$. 

\par
\medskip
{\bf Proof.}
Since the MC channel $G(s)$ is the circulant matrix, $G(s)$ can be diagonalized by the discrete Fourier transform (DFT) matrix $F\in\mathbb{C}^{n\times n}$ as 
\begin{equation}
    \Lambda(s) := F^{*}G(s)F,
\end{equation}
where $\Lambda(s)$ is the diagonal matrix and
\begin{equation}
    F = \frac{1}{\sqrt{n}}\left[\begin{matrix}
        1&1&1&\cdots&1\\1&\alpha&\alpha^2&\cdots&\alpha^{n-1} \\1&\alpha^{2}&\alpha^4&\cdots&\alpha^{2(n-1)} \\\vdots&\vdots&\vdots&\ddots&\vdots \\1&\alpha^{(n-1)}&\alpha^{2(n-1)}&\cdots&\alpha^{(n-1)(n-1)}
    \end{matrix} \right],\label{eq:DCT}
\end{equation}
with $\alpha = e^{-j2\pi/n}$. 
Note that $FF^*=I$ and $F^{-1}=F^*$, where $F^*$ is the conjugate transpose of $F$, since $F$ is the unitary matrix. Using the DFT matrix $F$ the characteristic polynomial (\ref{eq:character}) can be transformed as
\begin{eqnarray}
    p(s) &=& \det\left(F(I-h(s)F^{*}G(s)F)F^{*}\right)\nonumber\\
    &=& \det(I-h(s)\Lambda(s))\nonumber\\
    &=& \prod_i^n[1-h(s)\lambda_i(s)] = \prod_i^n\hat{p}_i(s).\label{eq:charatrans}
\end{eqnarray}
The theorem holds since Eq. (\ref{eq:charatrans}) shows that the roots of the characteristic equation $p(s)=0$ coincide with those of $\hat{p}_i(s)=0$ for all $i$. 

\par
\smallskip
Theorem 1 shows that $n$-dimensional MIMO system (\ref{eq:mimo}) with the circulant matrix $G(s)$ can be decomposed into $n$ SISO systems as shown in Fig. \ref{fig:mimo} (B), which facilitates the analysis for the stability of the MC system using methods such as Nyquist stability criterion developed in control engineering. 

\section{Numerical example}
\label{sec:numerical}

In this section, we demonstrate the proposed method to analyze the stability of a specific multi-agent MC system. In particular, we use Theorem 1 to decompose the MIMO MC system into the multiple SISO systems, and then analyze the stability of the decomposed SISO systems using Nyquist stability criterion to show the stable and unstable cases.

\par
\smallskip
We consider the MC system illustrated in Fig. \ref{fig:circ}, where each nanorobot has an activator-repressor-diffuser (ARD) genetic circuit \cite{yutaka2015coordinated}. 
The function $\bm{f}(\bm{x})$ of the reaction system is
\begin{equation}
    \bm{f}(\bm{x}) = \left[\begin{matrix}\displaystyle
        -\delta_{a}x_a + \gamma_{a}\frac{x_a^2}{K_{a}^2 + x_a^2}\frac{K_{r}^2}{K_{r}^2 + x_r^2}\\\displaystyle
        -\delta_{r}x_r + \gamma_{r}\frac{x_a^2}{K_{a}^2 + x_a^2}\frac{K_{d}^2}{K_{d}^2 + x_d^2}\\\displaystyle
        -\delta_{d}x_d + \gamma_{d}\frac{K_{r}^2}{K_{r}^2 + x_r^2} + k(x_e(t) - x_d(t))\\
        k(x_d(t) - x_e(t))
    \end{matrix} \right],\label{eq:f}
\end{equation}
where $x_a$, $x_r$, $x_d$, and $x_e$ are the concentrations of activator, repressor, signal molecule in the nanorobot, and signal molecule outside of the nanorobot, respectively. $\delta_i$ and $\gamma_i$ are the degradation rate and the production rate of the corresponding molecular species, and $K_i$ is the Michaelis Menten constant. The parameters values are referred from Case A of Sec. $\mathrm{I}\hspace{-1.2pt}\mathrm{I}
\hspace{-1.2pt}\mathrm{I}$ in \cite{yutaka2015coordinated} with $k=0.05\,\si{min^{-1}}$. 
Each MC channel $\Gamma_i$ has the same distance $L=50\,\si{\micro m}$ and the diffusion coefficient $\mu=83\,\si{\micro m^2\cdot s^{-1}}$.
In what follows, we approximate the MC system by the unit MC system with 4 nanorobots, and analyze the stability of the unit MC system around the equilibrium point based on the transfer functions using the proposed method.

\par
\smallskip
The transfer function matrix of the MC channel $G(s)$ is obtained as the $4\times 4$ circulant matrix shown in Eq. (\ref{eq:circulant}), and the reaction system is $H(s)=h(s)I$, where $h(s)=\bm{C}(sI-A)^{-1}\bm{B}$ with $A$ being the Jacobian matrix of Eq. (\ref{eq:f}). 
Using Theorem 1 the characteristic polynomial $p(s)$ of the MC system can be decomposed into $\hat{p}_i(s)$ with $n=4$, where $\hat{p}_1(s)=\hat{p}_4(s)$ and $\hat{p}_2(s)=\hat{p}_3(s)$. 

\par
\begin{figure}
    \centering
    \includegraphics[width=0.99\linewidth]{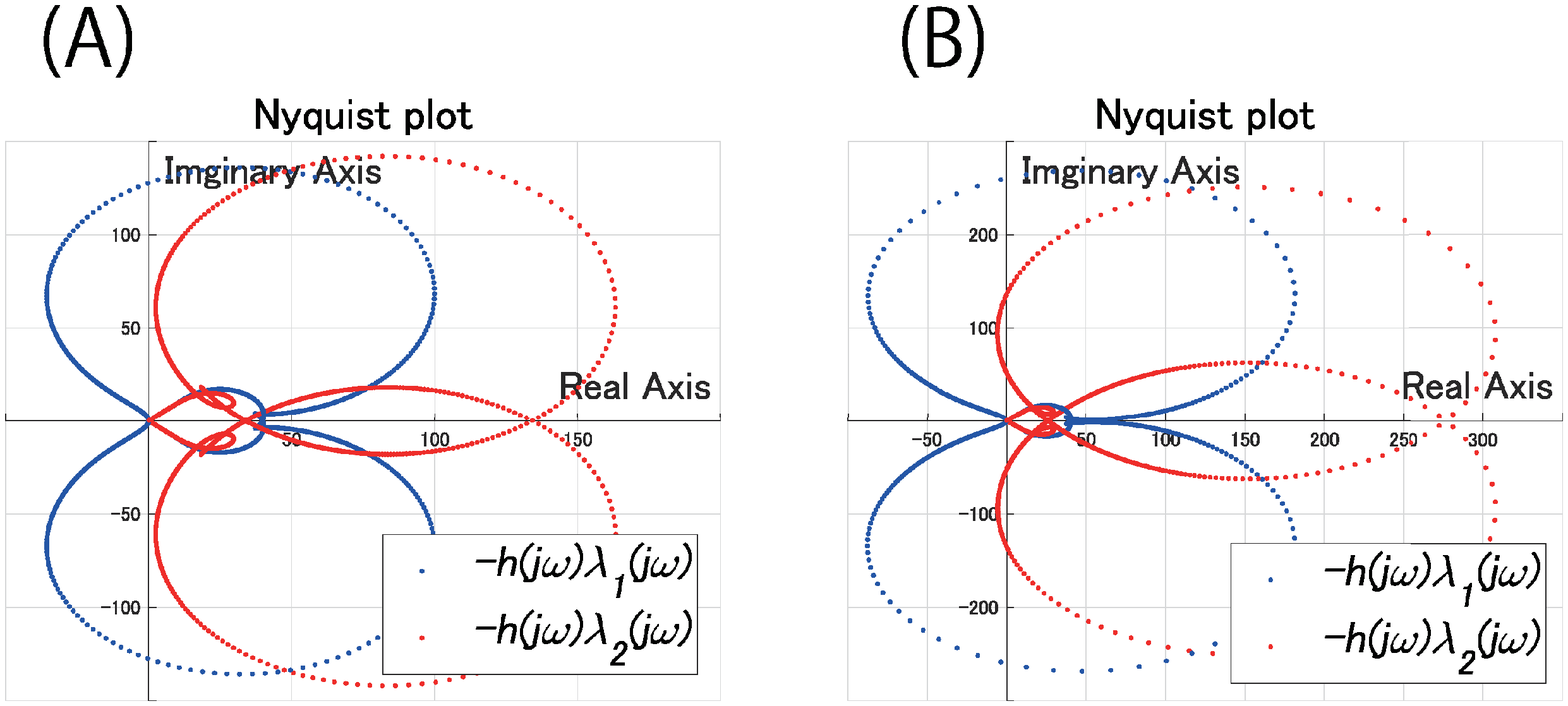}
    \caption{Nyquist plots of the MC system (A) with the production rate $\gamma_a=2.5\,\si{\micro M\cdot min^{-1}}$ and (B) with the production rate $\gamma_a=3.0\,\si{\micro M\cdot min^{-1}}$. }
    \label{fig:nyquist}
\end{figure}

\smallskip
We draw Nyquist plots to find the number of the roots $Z$ of $\hat{p}_i=0$ whose real parts are non-negative. The number $Z$ can be calculated by $Z=N+P$, where $N$ is the number that the trajectory of $-h(j\omega)\lambda_i(j\omega)$ encircles clockwise around the point $(-1,j0)$, and $P$ is the number of the non-negative roots of $-h(j\omega)\lambda_i(j\omega)$. Fig. \ref{fig:nyquist} (A) depicts the Nyquist plot of the MC channel for the production rate $\gamma_a=2.5\,\si{\micro M\cdot min^{-1}}$, which shows each trajectory of $-h(j\omega)\lambda_i(j\omega)$ does not encircle around $(-1,j0)$. Since $-h(j\omega)\lambda_i(j\omega)$ has no unstable poles, i.e. $P=0$, the number of the non-negative roots of the closed-loop is $Z=0$, and thus the system is stable. 
On the other hand, Fig. \ref{fig:nyquist} (B) depicts the Nyquist plot of the MC channel for $\gamma_a=3.0\,\si{\micro M\cdot min^{-1}}$, which shows each trajectory of $-h(j\omega)\lambda_i(j\omega)$ does not encircle around $(-1,j0)$. Since $P=2$, the number of the non-negative roots of the closed-loop is $Z=2$, which leads that the closed-loop of the MC system is unstable. 

\par
\begin{figure}
    \centering
    \includegraphics[width=0.99\linewidth]{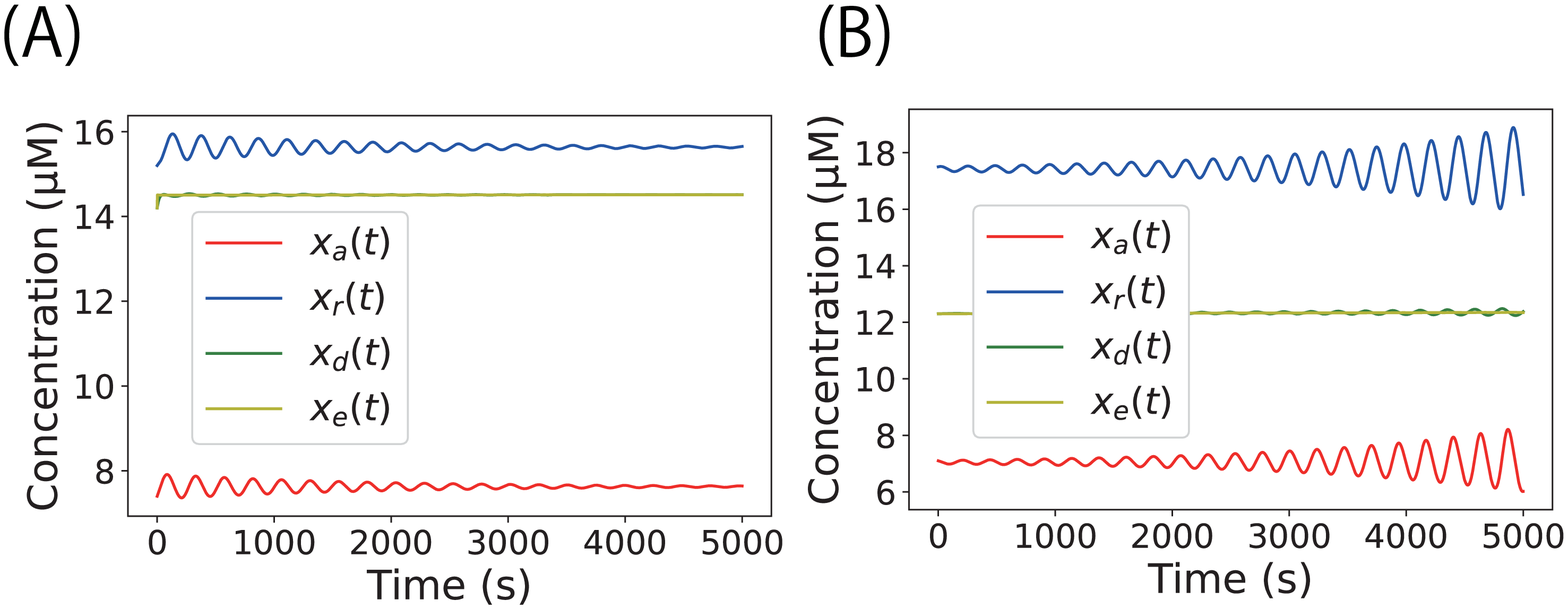}
    \caption{The behavior of each molecular concentration in the ARD genetic circuit around the equilibrium point. (A) The production rate $\gamma_a=2.5\,\si{\micro M\cdot min^{-1}}$ and the equilibrium point $\bm{x}^*=[7.6,15.6,14.5,14.5]$. (B) The production rate $\gamma_a=3.0\,\si{\micro M\cdot min^{-1}}$ and the equilibrium point $\bm{x}^*=[7.1,17.4,12.4,12.4]$.}
    \label{fig:response}
\end{figure}

\smallskip
Fig. \ref{fig:response} shows the concentration behavior of each molecular concentration in the ARD genetic circuit for different production rate $\gamma_a$ when perturbation inputs to each molecular concentration around the equilibrium point $\bm{x}^*$. The molecular concentration converges to the equilibrium point when $\gamma_a=2.5\,\si{\micro M\cdot min^{-1}}$, while the molecular concentration oscillates when $\gamma_a=3.0\,\si{\micro M\cdot min^{-1}}$. Thus, Fig. \ref{fig:response} verifies the proposed stability analysis method in Theorem 1, which is helpful for the analysis and design of multi-agent dynamical MC system.

\section{Conclusion}
\label{sec:conclusion}

In this paper, we have formulated the system theoretic model for multi-agent MC systems expressed by MIMO systems using the transfer functions based on a diffusion equation. We have then proposed the method to analyze the stability for the special case of the circulant structured multi-agent MC systems by decomposing the $n$-dimensional MIMO MC system into $n$ SISO systems. Finally, we have demonstrated the proposed method to analyze the stability of a specific MC system.


\end{document}